\UseRawInputEncoding
\documentclass{article}
\usepackage{authblk}
\usepackage{amsmath}
\usepackage{amsfonts}
\usepackage{amssymb}
\usepackage{graphicx}
\usepackage{amsthm}
\usepackage{blindtext}
\usepackage{hyperref}
\newtheorem{mydef}{Definition}
\newtheorem{mylem}{Lemma}
\newtheorem{myth}{Theorem}
\newtheorem*{myprob}{Problem}
\newtheorem{myconj}{Conjecture}
\input{Qcircuit}
\begin{document}
\title{ Parametrized 
Complexity of Quantum Inspired Algorithms}
\author[1]{Ebrahim Ardeshir-Larijani*}
\affil[1]{Institute for Research in Fundamental Sciences (IPM), Iran\\ e.a.larijani@ipm.ir}
\date{December 2021}
\maketitle
\tableofcontents
\section{Introduction}

Motivated by recent progress in quantum technologies and in particular quantum software, research and industrial communities have been trying to
discover new applications of quantum algorithms such as quantum optimization and machine learning. 
Regardless of which hardware platform these novel algorithms operate on, whether it is \emph{adiabatic} or \emph{gate based},  
from theoretical point of view, they are performing drastically better than their classical counterparts.

One of the remarkable examples is the quantum algorithm for solving linear systems of equations, introduced by Harow 
et. al.~\cite{hhl}. At its core, this algorithm uses the ``quantum phase estimation" procedure, which is used in many important quantum algorithms.
For the input matrix $A\in \mathbb{C}^{N\times N}$, where $N$ is dimension, and $Ax = b$, where $x,b\in \mathbb{C}^N$, and assuming that $A$, is $s$-sparse, that is at least $s$ entry at each row of the matrix is zero, the HHL algorithm terminates with time complexity $O(log(N)s^2\kappa^2/\epsilon)$. Here $\kappa$ and $\epsilon$ are condition number and approximation constant, respectively. The best classical algorithm for this problem has the complexity of 
$O(Ns\kappa log(1/\epsilon))$.
As we will detail briefly in the next section, the HHL algorithm has the following characteristics:
\begin{itemize}
\item It assumes that data, such as vector $b$, is \emph{already} loaded to a quantum state, such that quantum operations become possible on it. 
\item It assumes there is an efficient oracle that simulate Hamiltonians. 
\item As for solution, it produces approximation of a function of solution.
\end{itemize} 

Many quantum optimization and machine learning algorithms are based on performing matrix operations using quantum states and operation, in order to speed up data analysis. The main intuition behind Linear Algebra based quantum machine learning~\cite{qml}, is that data analytic deals with high dimensional vector spaces, where quantum computing can efficiently work with them. This is in contrast with many classical algorithms of matrix computation, where efficiency can be conditionally met, e.g. assuming \emph{low rank} and \emph{sparsity} of input matrices. 

In the Section~\ref{sec:LAQ}, we review quantum recommendation system~\cite{qrec} algorithm as one example of linear algebraic quantum machine learning and optimization algorithm. We review new remarkable result~\cite{qirecom} that how a classical algorithm can replace the quantum algorithm of recommendation system, only with poly logarithmic overhead. The main concept behind this algorithm is to drop the assumption of loading data into quantum states (as explained above for HHL). To this end we use a tailored data structure (QRAM), that can deal with both quantum and classical input, as well as a process called "dequantization", which produces a novel classical algorithms for problems such as recommendation systems and principal component analysis~\cite{qpca}. This process replaces quantum state preparation with ``\emph{sampling and query}" assumptions. The claim in~\cite{qrec} is that for the current applications, the latter assumptions are easier to satisfy than quantum state preparation. Formally, for a quantum protocol $P(\ket{\phi_1},\ldots,\ket{\phi_n})= \ket{\psi}$, dequantization means that we replace quantum state preparation with a classical sampling and query algorithm: $SQ(\phi_1,\ldots,\phi_n)=SQ(\psi)$.   

In Section~\ref{ssec:svt}, another line of research in terms of singular value transformation (SVT) is described. This is based on a recent algorithm for quantum SVT~\cite{qsvt}. Combining concepts of \cite{qirecom} and quantum SVT, we end up with a classical algorithm that under suitable sampling assumptions, computes SVT, independent of the dimension of the input matrices~\cite{smpldeq}.
This work generalizes the current trends of quantum algorithms of quantum linear algebra in terms of quantum SVT. 
Here the intuition is also built upon the fact that practical QML and optimization algorithms assume constraints on the input (e.g. sparsity) and generally   
loading input data into
an input quantum state and extracting amplitude data from an output quantum state are hard.

A fundamental goal of quantum-inspired algorithms is to reduce dimensionality of input matrices to speed up linear algebra computations.
Therefore, using matrix sketching techniques is natural here. The work~\cite{qinla} uses tools
from randomized numerical linear Algebra to obtain significant improvements in the sublinear terms for the dynamic (fast updates to the input matrix) and static settings. Its analysis relies on simulating \emph{leverage score sampling} and \emph{ridge leverage score sampling}, using a \emph{weight-balanced} data structure that samples matrices rows proportional to squared Euclidean norm ($l_2$ norm). This method will be reviewed in Section~\ref{ssec:nla}.  

From computational complexity point of view, the described quantum inspired algorithms are exponentially faster 
than their classical versions, given that the input satisfy some constraints. However, compared to their quantum counterparts, they suffer computational (polynomial) overhead. This raises the question, how these quantum inspired algorithms perform in practice? The work of Arrazola et. al.~\cite{qiapract}, tries to answer this question and concludes that the so called algorithms are in fact very promising in practical setting and benchmarks, by ensuring 
some stringent conditions are satisfied.

Studying the research mentioned above, motivates us to understand where the actual bottlenecks of the recent quantum inspired algorithms are? The complexity dependence to the input structure points us that perhaps the input size is not an appropriate measure of complexity, and hence we resort to \emph{Parametrized Complexity}~\cite{pc}. To this end we 
review some works in the area of parametrized complexity that we might find possible links to the above quantum inspired algorithms. In~\cite{ppca}, parametrized complexity of classical PCA is studied. Furthermore, we review the work about rigidity problem~\cite{rigid}. Moreover the works~\cite{fs,rigid} describe some recent work in the area of parametrized algorithm, relevant to machine learning and optimization. Finally, we review the parametrized complexity of \emph{matrix factorization}~\cite{pcfac}, a concept that is used in recommender system extensively.   
\section{Linear Algebraic Quantum Algorithms}\label{sec:LAQ}
In this section, first we review the HHL algorithm for solving linear systems of equation. Then we review the algorithm of 
Lloyd et. al.~\cite{qpca}, for the problem of principal component analysis (PCA), which is widely used in the area of 
machine learning. Finally, we review the Quantum Singular Value Transformation algorithm~\cite{qsvt}, which is a unification methods for quantum algorithms based on quantum phase estimation, in a single algorithm.
\subsection{Solving Linear systems of equations}\label{ssec:hhl}
A linear system of equation can be described by an equation of the form $Ax=b$, where $A$ is a matrix and $b$ and $x$ are vectors of finite dimension, e.g. $N$. In our case, $A\in \mathbb{C}^{N\times N}$ and $x,b\in \mathbb{C}^N$. The goal is to find (estimate) the solution, i.e. $x$. The algorithm~\cite{hhl}, uses a sub algorithm, like many other quantum algorithm, for phase estimation. For a description of this useful and important algorithm see~\cite{NC}. Here we only describe what the phase estimation does. 

Suppose $U\in \mathbb{C}^{2^m\times 2^m}$ be a unitary matrix and $\ket{\psi}_m\in \mathbb{C}^{2^m}$ be its  
eigenvectors with associated eigenvalue $e^{2\pi i \theta}$. The \emph{Quantum Phase Estimation (QPE)}, approximates $\theta$ for given unitary matrix. More formally:
\[QPE(U,\ket{0}_n,\ket{\psi}_m)= \ket{\hat{\theta}}_n\ket{\psi}_m\]
Where binary approximation of $2^n\theta$,  truncated to $n$ digits is denoted by $\hat{\theta}$.
In the case of HHL algorithm, the unitary $U$ has the form of $e^{iAt}$, where $A$ is the matrix coming from linear system of equations. Furthermore, we can rewrite $U$ as following:
\[e^{iAt} = \sum_{j=0}^{N-1} e^{i\lambda_j t}\ket{u_j}\bra{u_j}\]
Now, with eigenvector $\ket{u_j}_{n_b}$ and eigenvalue $e^{i\lambda_j t}$, the QPE outputs $\ket{\hat{\lambda_j}}_{n_l}\ket{u_j}_{n_b}$. Here the $\hat{\lambda_j}$ is $n_l$-bit binary approximation to $2^{n_l}\frac{\lambda_j t}{2\pi}$.

\begin{figure}
\begin{center}
\includegraphics[width = 10 cm, height = 5cm]{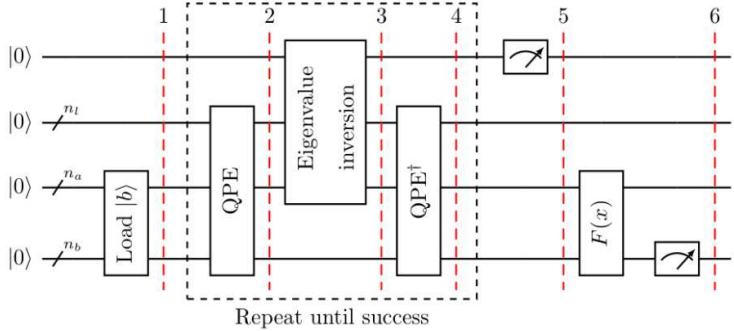} 
\end{center}
\caption{HHL Algorithm steps}
\label{fig:hhl}
\end{figure}

Now we can describe HHL algorithm for solving linear system of equation in six main steps. 
Schematically, these are shown in Figure~\ref{fig:hhl} , in the language of quantum circuits~\cite{NC}.
Here we have three quantum registers. One is used for storage of binary representation (indexed by $n_l$) eigenvalue of $A$.
We also a register to load vector $b$, indexed by $n_b$, and hence the dimension of the problem is $N=2^{n_b}$. 
Note that we have ancillary qubits in a third register, indexed by $n_a$.

The description of the algorithm is as following~\footnote{for a detailed description and also a numerical example and implementation please have look at:\url{https://qiskit.org/textbook/ch-applications/hhl_tutorial.html} }

\begin{enumerate}
\item Loading vector $b$ into the quantum state $\ket{b}\in \mathbb{C}^N$, i.e. applying the transformation: $\ket{0}_{n_b}\mapsto \ket{b}_{n_b}$.
\item Applying QPE to the unitary $U=e^{iAt}$. Now the quantum register has the state: 
\[\sum_{j=0}^{N-1} b_j\ket{\lambda_j}_{n_l}\ket{u_j}_{n_b}\]
here the $n_l$-bit binary representation of $\lambda_j$ is encoded by $\ket{\lambda_j}_{n_l}$.
\item Apply a conditional rotation on $\ket{\lambda_j}_{n_l}$ and add auxiliary qubit (first row in Figure~\ref{fig:hhl}).
\item Apply $QPR^{\dagger}$.
\item Measure auxiliary qubit in computational basis. If the outcome is $1$, then the register would be in post-measurement state:
\[\sqrt{\frac{1}{\sum_{j=0}^{N-1}|b_j|^2/|\lambda_j|^2}}\sum_{j=0}^{N-1}\frac{b_j}{\lambda_j}\ket{0}_{n_l}\ket{u_j}_{n_b}\]
which corresponds to the solution up to the normalization factor.
\item Apply an observable $M$ to calculate $F(x) = \bra{x}M\ket{x}$.
\end{enumerate}

\subsection{Quantum Recommender System}\label{ssec:qrecm}
In recommendation system, the goal is to use information on purchase/rating of $n$ products/services by a group of $m$ clients, to provide personalize recommendation to individual clients. This information is modeled by a $m\times n$ preference matrix, where the entries indicates how user $i$ evaluates $j$.  
The challenge is the preference matrix is not known a priori, such that by looking into the high evaluation make good recommendation to the users. The information from clients are received online. So the goal is to infer a good approximation of preference matrix that reflects high value rating. 

One important characteristics of preference matrix in recommender systems are that they are 
\emph{low rank}. This is because users interests fall in certain types, not all of the products.

The first algorithm by Kerenidis et. al.~\cite{qrec}, uses a technique to reconstruct the a low matrix approximation by using samples (available information) of preference matrix, up to a 
measure of matrices, e.g. $l_2$ or Frobenius norms.

The result of sampling technique and matrix reconstruction of quantum recommender systems is 
an online computation that requires $O(poly(k)polylog(mn))$, where $k$ is the rank of the input matrix.

One important assumption of \cite{qrec} is that there is an efficient algorithm and data structure to load classical matrix into quantum states. 
Theorem~\ref{th:1} shows with the assumption
that the transformation $A\mapsto \ket{A}$ can be done efficiently,
we can store the input of quantum recommendation system optimally. However,
this assumption is dropped in \cite{qirecom}. More details on the algorithm and in particular sampling technique can be found in \cite{qrec}.
We will detail some aspects of sampling    
in the next section.   

\begin{myth}\cite{qrec}
\label{th:1}
Let $A\in \mathbb{R}^{m\times n} $ be a matrix. The input of the recommendation system with the format $(i,j,A_{ij})$ coming in arbitrary order. Let $w$ be the number of input that already been taken. Then, there exists a data structure to store the input of the recommendation system with the following features:
\begin{enumerate}
    \item The overall size of the data structure is $O(w.log^2(mn))$.
     \item The processing time for storing new input is $O(log^2(mn))$.
     \item Let $\tilde{A}\in \mathbb{R}^m $ defined by $\tilde{A_i}= ||A_i||$. A quantum algorithm with access to this data structure can perform the following mappings in time $polylog(mn) $: 
     $\tilde{U}: \ket{i}\ket{0} \mapsto \ket{i}\ket{A_i} $ for $i\in [m]$, and $\tilde{V}: \ket{0}\ket{j}\mapsto \ket{\tilde{A}} \ket{j}$ for $j\in [m]$ (projecting $A$ rows and columns, respectively). 
\end{enumerate}
\end{myth}

\subsection{ًQuantum principal component analysis}\label{ssec:qpca}
One of the widely used dimension reduction techniques for large data sets is  
principal component analysis (PCA).
In PCA we use spectral decomposition of matrices in terms of eigenvectors and corresponding eigenvalues, and discards eigenvalues below certain threshold, to obtain 
a good low rank approximation of the input matrix. 

Let data are represented by $d$-dimensional vectors $v_j$. Then the covariance matrix $C$ illustrate the correlation between different components of the data
and is defined as $C = \sum_j v_jv_j^T$. Then PCA diagonalize the covariance matrix:
\[C = \sum_k e_k c_k c_k^{\dagger} \]
Now only a few eigenvalues $e_k$ are non-zeor or sufficiently large and the rest could be discarded. The remaining called principal components and any given data vector $v$,
can be succinctly described by these $k$ components: $v=\sum_k e_k c_k$.
This operation requires $O(d^2)$ in terms of time and query complexity. 

The quantum algorithm for PCA~\cite{qpca}, in the first step needs to map a randomly chosen classical data 
to a quantum state $v_j\mapsto \ket{v_j}$. This is usually done with a dedicated data structure,called QRAM. Because these data are picked in random, the resulting state has the density matrix $1/N \sum_j \ket{v_j}\bra{v_j}$, where $N$ is the size of the data set. This is in fact equals to covariance matrix up to a constant factor. Now by incorporating a sampling technique~\cite{qpca} and phase estimation (as in section~\ref{ssec:hhl}) we can find estimation of eigenvalues and corresponding eigenvectors:
\[\ket{v}\mapsto \sum_k v_k \ket{c_k}\ket{\tilde{e_k}}\]
Now the properties of principal components of $C$ can be inferred from measurements on the quantum representation of its eigenvectors. The time and query complexity of this algorithm is
$O((log N)^2)$. 
\subsection{Quantum singular value transformation}\label{ssec:qsvt}
The area of quantum algorithms involves with two types of algorithms: QPE based (such as the algorithms described in the previous sections) and Quantum Fourier Transformation algorithms~\cite{NC} (such as Shor factorization). The aim of 
Quantum singular value transformation (QSVT)~\cite{qsvt}, is to find a general framework for
QPE based algorithms.

Many important applications of quantum computing use spectral decomposition and transformations, and therefore use QPE, which we reviewed some of them in the previous sections.
Nevertheless, these algorithms work with quantum states and need to perform encoding and then reading measurement outcomes to learn certain properties up to a specified precision $\epsilon$.
Therefore the improvements in terms of computational complexity of these algorithms is merely
\emph{polynomially}. Of course for large scale optimization applications, such as semi-definite quantum program solvers this polynomial improvement matters significantly~\cite{qsdp}.

The central concept of QSVT is \emph{projected unitary encoding}. Let $\Pi$ and $\tilde{\Pi}$ are orthonormal projectors. Let $U$ be any unitary. 
For an operator $A$, we say $U$, $\Pi$ and $\tilde{\Pi}$ form a projected unitary encoding, if we have: $A = \tilde{\Pi} U \Pi$.

Let $P\in\mathbb{C}[x]$ be an odd polynomial and $A = W\Sigma V^{\dagger}$ be a singular value decomposition (SVD). Now the singular value transformation is defined by:
\[P^{SV}(A):= W P(\Sigma)V^{\dagger} \]

The main result of \cite{qsvt} is stated as following: 

For any odd polynomial of 
degree $d$, i.e. $P\in\mathbb{C}[x]$, where $|P(x)|\leq 1$, there is a unitary 
$U_{qsvt}$ which can be implemented by a quantum circuit consists of only $U$ and $U^{-1}$, $d$ times throughout, such that:

\[A = \tilde{\Pi} U \Pi \Rightarrow P^{SV}(A)= \tilde{\Pi} U_{qsvt} \Pi \] 

See~\cite{qsvt} for full details as well as examples.

\section{Quantum-Inspired Algorithms for Matrix Computation}\label{sec:QIQ}
In this section we review the new trend in designing quantum inspired algorithms, appeared in a series of publications~
\cite{qirecom,qireg,qipca}. We also review the singular value transformation based technique~\cite{smpldeq}.
In both frameworks, we deal with dequantization concept, as described in introduction section.
Furthermore, we explore new techniques from ~\cite{qinla}, where more advanced sampling methods for matrices such as leverage score, is used to achieve improvement. Also new classical data structure to support dequantization is introduced in this work.   
Our review captures the main ideas and drop most of technical details for brevity. 
\subsection{Sample and Query Techniques}
In this section we explain on of the central notions introduced in~\cite{qirecom} and the subsequent works. 
The goal is to construct a \textit{classical} input model which is close quantum input model (quantum state), while we maintain complexity of sublinear time. To do so we define \textit{sampling and query} access as following:
\begin{mydef}
We define query access model for $v\in \mathbb{C}^n$, denoted by $Q(v)$ and sample and query model, denoted by $SQ(v)$:
\begin{enumerate}
\item for all $i\in [n]$,
we have oracle access and can query for $v(i)$. Similarly, for a matrix
$A\in \mathbb{C}^{m\times n}$, we have $Q(A)$ if for all $(i,j)\in [m]\times[n]$,
we can query for $A(i,j)$.
\item We have sampling and query access, $SQ(v)$ if following conditions are met:
\begin{enumerate}
\item For any $v$, we can query $Q(v)$.
\item If we can get independent samples, $i$, over $[n]$ with the distribution $D_v\in \mathbb{R}^n$, defined as
$D_v := |v(i)^2|/\Vert v \Vert^2$.
\item We can query for $\Vert v \Vert$.    
\end{enumerate}
\end{enumerate}
\end{mydef}
Note that in order to implement the above input model, we need a dedicated (dynamic) data structure. In a typical
Random Access Memory (RAM) model, query access has linear computational complexity. However, we are interested in 
a (classical) data structure that support sublinear access complexity. In~\cite{qirecom} uses a data structure that \cite{qrec} uses for preparation of quantum states, which satisfies all condition we need. 

\begin{figure}[h]
\begin{center}
\includegraphics[width = 10 cm, height =5 cm]{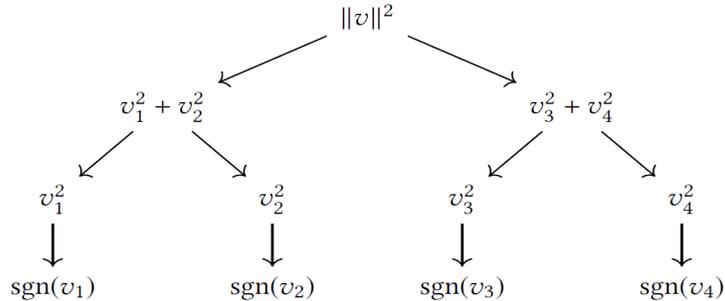} 
\end{center}
\caption{Dynamic binary search tree for storing inputs of quantum inspired algorithms}
\label{fig:dbst}
\end{figure}

One of the dynamic data structures which is used by mentioned QML algorithms is a binary search tree, that 
leaf nodes store $v_i$ using the weight $v_i^2$ and $sgn(v_i)$. The interior nodes weight is the sum of their children weights. For updating and entry, all nodes above the corresponding leaf has to be updated.
For sampling, start from the root of tree and randomly choose a child, where the probability is proportional to its weight. Note that because we mostly deal with sparse input (such as recommendation systems), we only include non zero nodes in the tree. Figure~\ref{fig:dbst}, shows and example of such tree for a vector $v\in \mathbb{R}^4$.  

Now we have the following lemma:

\begin{mylem}\label{lem:1}
To store a vector $v\in \mathbb{R}^n$ with $w$ nonzero entries, there exists a data structure, supporting the following
operations:
\begin{enumerate}
\item Querying and updating an entry of $v$ in $O(log n)$ time.
\item Querying $\Vert v \Vert^2$ in $O(1)$ time.
\item Sampling from $D_v$ in $O(log n)$ time.
\end{enumerate}  
\end{mylem} 

Note that the actual format of the input in our problems is matrix so for a matrix 
$A\in \mathbb{R}^{m\times n}$, we form a vector $\tilde{A}\in \mathbb{R}^m$, where the $i^{th.}$ entry equals to 
$\Vert A_i \Vert$ ($l_2$ norm of the $i^{th.}$ row of $A$). Then we have the following lemma~\cite{qirecom}
\begin{mylem}\label{lem:2}
There exists a data structure storing a matrix $A\in \mathbb{R}^{m\times n}$ with $w$ nonzero entries in $O(w log mn)$ space,
supporting the following operations:
\begin{itemize}
\item Reading and updating an entry of $A$ in $O(log mn)$ time;
\item Finding $\tilde{A}_i$ in $O(log m)$ time;
\item Finding $(\Vert A \Vert_F)^2 $ (Frobenius norm) in $O(1)$ time;
\item Sampling from $D_{\tilde{A}} $ and $D_{A_i}$ in $O(log mn)$ time.
\end{itemize}
\end{mylem}

\begin{figure}[h]
\begin{center}
\includegraphics[height = 4 cm , width = 11 cm]{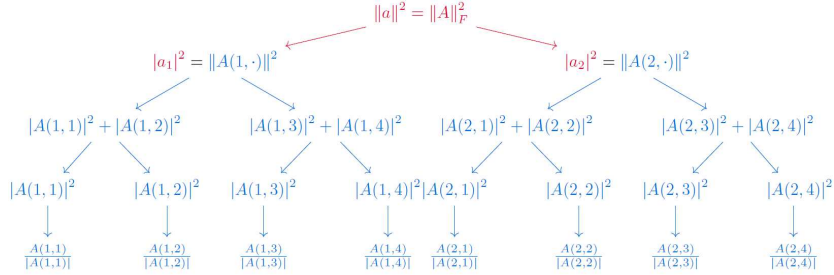} 
\end{center}
\caption{Dynamic data structure for storing $A\in \mathbb{R}^{2\times 4}$}
\label{fig:ddm}
\end{figure}

By having a copy of the data structure specified in Lemma~\ref{lem:1} for each
row of $A$ and $\tilde{A}$ we can form the above data structure. This has
all of the desired properties, and in fact, is the data structure Kerenidis and Prakash~\cite{qrec} use
to prepare arbitrary quantum states, which enables our algorithm to
operate on the same input. Figure~\ref{fig:ddm} shows an example of such data structure for a matrix 
$A\in \mathbb{R}^{2\times 4}$.

\begin{myth}\label{th:1}
[\cite{qirecom}] Let $A$ be an input matrix that supports query and sampling as specified in Lemma~\ref{lem:2}, 
suppose $i\in [m]$ be a row and $\sigma, \eta$ be singular value thresholds for low rank approximation of $A$ (denoted by $A_{\sigma,\eta}$). For a small $\epsilon$, there is a classical algorithm that has an output distribution $\epsilon$-close to $D_{M_i}$, where $M\in \mathbb{R}^{m\times n}$, satisfying 
$\Vert M - A_{\sigma,\eta} \Vert_F \leq \epsilon \Vert A \Vert_F$, in query and time complexity

\[O (poly( \frac{\Vert A \Vert_F}{\sigma},\frac{1}{\epsilon},\frac{1}{\eta},\frac{\Vert A_i\Vert}{\Vert D_i\Vert} ) )  \]    
\end{myth}

The Theorem~\ref{th:1} states its algorithm has a runtime indpendent of the input dimension ($m,n$). To implement the needed sampling operations we need the dynamic data structure of Lemma~\ref{lem:2}. Note that using that data structure will incur additional $O(log (mn))$ overhead. Comparing this algorithm and the quantum algorithm of ~\cite{qrec}, shows that in the quantum version of producing low rank approximation, the dependence on $\epsilon$ is logarithmic, compare to the classical quantum inspired version. 

Now by using the above theorem and impose the constraint that linear algebraic operations should not add the cost of reading a full row or column of the input, and by using a number of sampling routines (see~\cite{qirecom} for full details)
and applying it to the input of recommendation system, i.e. preference matrix, we have the following theorem:

\begin{myth}\label{th:2}
[\cite{qirecom}] By applying Theorem~\ref{th:1} to the input model of recommendation system, which uses the dynamic data structure in~\cite{qrec}, we obtain the same complexity bounds up to constant factors for small $\epsilon$.
\end{myth}

For the PCA problem we use similar techniques of sampling and query. The original algorithm of Lloyd et. al.~\cite{qpca} best performs in low rank setting therefore we have the following problem:

\begin{myprob}
[PCA for low-rank matrices] For a Given a matrix $A\in \mathbb{C}^{m\times n}$, where we have access to $SQ(A)$ such that $AA^{\dagger}$ has top $k$ eigenvalues $\{\lambda_i\}_{i=1}^{k}$
and eigenvectors $\{v_i\}_{i=1}^{k}$
, with probability $\geq 1-\delta$, compute eigenvalue
estimates $\{\hat{\lambda_i}\}_{i=1}^{k}$
such that
$\sum_{i=1}^k |\hat{\lambda_i}-\lambda_i|\leq \epsilon Tr (AA^{\dagger}) $
and eigenvectors $\{\hat{v_i}\}_{i=1}^{k}$, (with access to $\{SQ(\hat{v_i})\}_{i=1}^{k}$)
such that
$\forall i, \Vert v_i - \hat{v_i} \Vert \leq \epsilon $.
\end{myprob}

The quantum inspired algorithm in $\cite{qipca}$ exactly solve this problem by using sampling techniques and low rank approximation. For details of its algorithm see ~\cite{qipca}, page 3.

\subsection{Quantum inspired Singular Value Transformation (SVT)}\label{ssec:svt}
As we described in Section~\ref{ssec:qsvt}, the quantum algorithms based on phase estimation can be generalized by transformation on singular values.

In fact transforming singular values is the crux of many machine learning
methods. 
For instance,  
QML algorithms such as quantum support vector machines \cite{qsvm},
principal component analysis \cite{qpca}, training Boltzmann machines for state tomography~\cite{bm},
all are based on singular value transformation.

Note that the speed up in quantum algorithms stems from the ability of unitary operations in describing evolution of quantum systems with exponential size.
This leads to the idea of block encoding of large matrices which is a special case of projected encoding described in Section~\ref{ssec:qsvt}.

However, the aim of this section is to extract a sampling based methods~\cite{smpldeq} for the seminal work of quantum singular transformation of ~\cite{qsvt}. 

The main result of~\cite{smpldeq} states that the for a matrix $A\in \mathbb{C}^{m\times n}$,
with access to $ًُSQ(A)$, we can obtain a succinct characterization of singular transformation of $A$, efficiently. 

\begin{mydef}[Even singular value]
For a given $A\in \mathbb{C}^{m\times n}$ and a function $f:[0,\infty)\mapsto \mathbb{C}$,
and SVD of $A = \sum \sigma_i u_i v_i^{\dagger} $, the even singular value transformation 
of $A$, defined as $f(\sqrt{AA^{\dagger}})= \sum f(\sigma_i)v_iv_i^{\dagger}$
\end{mydef}

\begin{myth}
Assume for a the input $A\in \mathbb{C}^{m\times n}$ we have access to $ًُSQ(A)$
Let $f:[0,\infty)\mapsto \mathbb{C}$ such that $f$ is $L$-\textit{Lipschitz} and
$\bar{f}(x) = (f(x) - f(0))/x$.
Then for small enough $\epsilon,\delta\geq 0$, we can find a subset of $A$'rows,
denoted by $R\in \mathbb{C}^{r\times n}$ and a subset of $R$'s column, denoted by 
$C\in \mathbb{C}^{r\times c} $ such that:
\[Pr \,[\Vert R^{\dagger} \bar{f}(CC^{\dagger})R  + f(0)I - f(A^{\dagger}A)    \Vert\, >\epsilon ]< \delta  \]
If sampling and query takes $T$ time, then finding $R$, $C$ and computing $\bar{f}(CC^{\dagger})$ has the complexity $O(r^2c + rcT)$.
\end{myth} 

The $R^{\dagger} \bar{f}(CC^{\dagger})R$ in \cite{smpldeq} is called \emph{RUR} decomposition.
In that case the above theorem says that RUR decomposition expresses a good approximation of
input matrix as a linear combination of $r^2$ outer products of its rows.
See \cite{smpldeq} for thorough details.  
\subsection{Methods from Randomized Numerical Algebra}\label{ssec:nla}
In the previous sections we drew parallels to quantum algorithms by implementing the so called dequantization technique.
One can observe the heavy reliance of this technique on randomized numerical linear algebra, where low matrix rank computation and matrix sketching has been used at large. In a recent work~\cite{qinla} raises the following questions:
Is the running time of quantum inspired algorithms, that are based on sample and query techniques, and have sublinear terms is improvable? Here the goal is to maintain the efficient creation of the dynamic data structure in $O(nnz(A))$ of a given input matrix $A$, and updating it in $O(log~n)$. Another question is can we use a vast number of techniques in randomized numerical linear algebra in this context?

The main contribution of ~\cite{qinla} is answering to the above questions. Given the technicality of the methods in the context of numerical linear algebra we refer the reader to the main text and we only give statement of the improvement result for the quantum inspired algorithm for recommender systems here.

Sketching is a widely used solution for reducing dimensionality of matrix computation. The goal is to find a new matrix with its rows (or columns) are subset of the target matrix but still hold its main structures. Lemma 9 of ~\cite{qinla}, summarizes the main sketching techniques, relevant to this work.


For a given $n\times d$ matrix $A$ of rank $k$, the lower bound for the singular value $\sigma_{k}(A)$, is denoted by $\hat{\sigma_k}$. Let $\psi_{\lambda} = \frac{\Vert A \Vert^2_F}{\lambda + \hat{\sigma_k}^2}$. Let $\psi_k$ be the estimate of $\frac{||A||_F^2}{\sigma_k(A)^2}$.
Now Theorem~\ref{th:sk}, presents a sampling technique that compared to ~\cite{qirecom}, has a lower overhead and smaller relative error. For further details on how a dynamic data structure can be constructed from sketching matrix, see ~\cite{qinla}.


\begin{myth}\cite{qinla}
Let $A\in \mathbb{R}^{n\times d}$ is represented by a sampling data structure. Let $k$ and $\epsilon$, be target rank and error parameter, respectively. Suppose $A_k$ be the best $k$ rank approximation of $A$. Then we can find sampling matrices $S$, $R$ and $W$ (a rank $k$ matrix) such that:
\[||ARWSA - A ||_F \leq (1+ O(\epsilon)) ||A - A_k ||_F \]
with the running time:
\[\tilde{O}(\epsilon^{-6}k^3+\epsilon^{-4}\psi_{\lambda} (\psi_{\lambda}) + k^2+k\psi_k   ) \]
\label{th:sk}
\end{myth}

\section{Complexity Analysis of Quantum - Inspired Algorithms}
In this section we complement theoretical analysis of quantum inspired algorithms that are based on sampling and query and matrix sketching techniques, by reviewing implementation and numerical benchmarking, first reported in ~\cite{qiapract}.
The aim of such experiment is to acquire an understanding of the \emph{practical complexity} of quantum inspired algorithms 
in terms of \emph{erroneous} and \emph{runtime}. The main finding is that although we have seen in the previous sections a hefty polynomial overhead compare to quantum algorithms, however in practice and in the case we only have indirect access to massive data sets via query, quantum inspired algorithms performs very well. This is a reminding to the fact that computational complexity of these problems does not depend on the dimension of input, but other quality of the input and the manner input is accessible. In the following we briefly report on the findings of ~\cite{qiapract}.

\begin{figure}[h]
\begin{center}
\includegraphics[width = 7 cm , height = 5 cm]{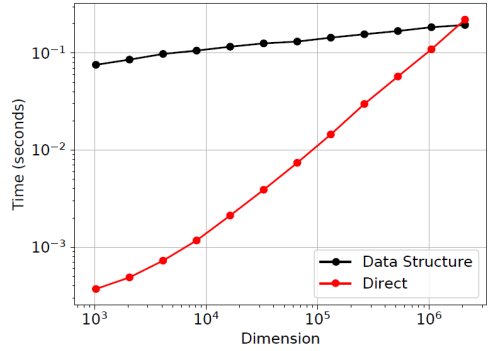} 
\end{center}
\caption{Runtimes for generating one thousand samples from a length-square distribution}
\label{fig:l2norm}
\end{figure}

The first example concerns with sampling according to the $l_2$ norm. For the dimension of less than $10^6$, using direct sampling is faster than using dynamic data structure of \cite{qrec} and \cite{qirecom}. The Figure~\ref{fig:l2norm}
illustrate the runtime of producing $1000$ sample of $l_2$ norm distribution according to the dynamic data structure as explained previously and direct sampling method.

The second example which we mention here (see more case studies and details in~\cite{qiapract})
is movie recommendation systems based on MovieLens 100K data set. 
It consists of a preference matrix with 100,000 ratings from 611 users across
9,724 movies. 
where ratings are scaled within the range $[0.5, 5]$.
Here majority of users watch only a small fraction of available movies, and therefore the matrix is sparse. 
The preference matrix has full rank of $k = 611$, its largest and smallest singular values are respectively
$\sigma_{max} = 534.4$, $\sigma_{min} = 2.95$, 
and its condition number (for $l_2$ norm $\frac{\sigma_{max}}{\sigma_{min}}$) is $\kappa = 181.2$.
The algorithm tries to predict missing ratings and subsequently recommend movies
that have a high predicted rating, which we explained in Section~\ref{sec:QIQ}.

\begin{figure}[h]
\begin{center}
\includegraphics[width = 12 cm, height = 2 cm]{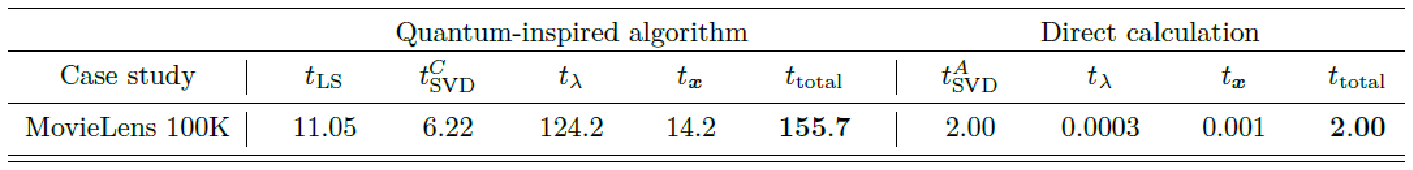} 
\end{center}
\caption{Running times (in seconds) for the quantum-inspired algorithm for recommendation systems. 
The parameter $t_{LS}$ is the time required to construct
length-squared (LS) probability distributions over rows and columns}
\label{fig:movie}
\end{figure}

Figure~\ref{fig:movie}~\cite{qiapract} compares the running times between the quantum-inspired algorithm and the direct
calculation method. 

For small size matrices
direct computation of SVD  can be done in a shorter time. 
The runtime of the quantum-inspired algorithm
is dominated by the coefficient estimation step, even when the resulting errors in estimating
these coefficient is relatively high.
As a conclusion,
the numerical examples we mentioned here suggest that when applied to moderately-sized
 data sets, because they rely on a more intricate procedure, the quantum-inspired algorithms
take more time than exact diagonalization, and because they rely on sampling for coefficient
estimation, lead to higher inaccuracies. These results show that in order to provide a speedup
over preexisting classical algorithms, the quantum-inspired algorithms must be applied to 
\emph{extremely large} data sets where exact diagonalization is impossible and where even the linear scaling of SVD approximation algorithms (for example those based on monte carlo methods such as FKV~\cite{fkv}) 
prevents its direct application.


It is also relevant to examine the complexity of quantum inspired algorithm where we have perturbation in the input. This has be done in the context of smooth analysis in 
a recent work of Dunjko et. al.~\cite{smooth}. 
They proved, using smooth complexity that 
if the quantum (inspired) algorithm is robust against small entry-wise input perturbation, state preparation can always
be achieved with constant queries.

The entry-wise access is done by an oracle $O_x$, with an arbitrary function $f(x)$:
\[\sum_{i,j}]\alpha_{i,j}\ket{i}\ket{j}\xrightarrow{O_{f(x)}}
\sum_{i,j}]\alpha_{i,j}\ket{i}\ket{j+f(x_i)}
 \]

For simplicity we assume $f(x_i)=x_i$. Note that in the classical setting, random access memory, provides the mapping $i\mapsto f(x_i)$. Now generating $l_2$ samples from entry-wise oracle and unstructured search has lower bound $\Omega (D)$ queries for $x\in \mathbb{R}^D$.
This can be achieved by QRAM model~\footnote{The paper~\cite{smooth} techniques can be applied to both quantum and quantum inspired algorithms. Here we mention results for quantum inspired setting}.

To model noisy data input, it is assumed that 
Gaussian element wise
perturbation occurs. Subsequently, smooth complexity is defined.

\begin{mydef}
[Smooth Complexity~\cite{smooth}] 
Given an algorithm $A$ with an input domain 
$\Omega_D = \mathbb{R}^D$, the smoothed complexity of $A $with $\sigma$-Gaussian
perturbation is defined as:
\[max_{x\in [-1,1]^D} \mathbb{E}_g[T_A(x+g)]\]
\end{mydef}

Here $g$ is a Gaussian random vector with variance $\sigma^2$, and
$T_A$ denotes the runtime of $A$. Having defined smooth complexity, the main result of \cite{smooth} is the following theorem:

\begin{myth}
Given oracle access, $O_x$ to the entries of 
$x\in \mathbb{R}^D$, the amplitude encoding of $x$ into $\ket{x}$ (state preparation) has smoothed complexity $O(1/\sigma)$.
\end{myth}

\subsection{Parametrized Complexity}
In this section we review some works in the context of classical algorithms that may have relevance to parametrized complexity analysis. The goal of our project is to extend parametrized complexity concepts to the quantum inspired algorithms which we explained in the previous sections.

A central concept in parametrized complexity~\cite{pc} is \emph{Fixed Parameter Tractability (FPT)}.
For FPT algorithm, the performance is heavily depends on some \emph{fixed parameters} rather than the \emph{input size}.
So we can view these problems in sliced way where each slice corresponds to a tractable problem. 
Another important concept is "kernelization"~\cite{pc}, where we reduce the parametrized input to a new parametrization where the algorithm performs better. The formal definition of FPT is as following:

\begin{mydef}\label{def:fpt}
[Fixed parameter tractable]
We say an algorithm $A$ is FPT if for all parametrized input $\langle x,k \rangle$, constant $c$, and a commutable function
$f$, $A$ has a running time $f(k)|x|^c$. Here $c$ is independent from $|x|$ and $k$. 
\end{mydef}

Parametrized algorithms with higher complexity belong to $XP$, where the complexity bound is in terms of commutable function $f$ and $g$ : $f(k)|x|^{g(k)}$.

The first example concerns with recommendation system (which we explained its quantum version in Section~\ref{ssec:qrecm}). Here~(\cite{pcfac}) the problem formulated as ``\emph{weighted low rank approximation}" and 
``matrix completion problems". The formal definition of these problems are as following:
\begin{mydef}\label{th:wlra}
[Weighted low rank approximation]
\[\textit{Input:}~A\in \mathbb{R}^{n\times n},W\in \mathbb{R}^{n\times n}, k\in \mathbb{N}, \epsilon > 0\]
\[\textit{Output:}~ U,V^T \in \mathbb{R}^{n\times k} s.t. \Vert W\circ (UV - A) \Vert_F \leq (1+\epsilon) OPT  \]
Where $OPT := Min_ A' \Vert W\circ (A' - A) \Vert_F, rank(A')\leq k$
\end{mydef}
 
In the matrix completion problem, we are given a matrix with missing entries
and the task is to complete the matrix in a way
that 
(1) minimizes the number of distinct rows and columns, 
(2) minimizes the number of distinct column, or 
(1) minimizes the rank.
It is shown in \cite{mcp} that this problem is FPT for bounded domain.  

\begin{mydef}[matrix completion problem]
\[\textit{Input:}~ A\in \{\mathbb{R},?\},~ \textit{known entries}~\Omega \in [n]\times [n], k\in \mathbb{N} \]
\[\textit{Output:}~ A_{\bar{\Omega}}~ s.t.~ rank(A) = k\]
\end{mydef}

Now the \cite{pcfac}, considers the weighted low rank approximation with these three parametrizations.

\begin{myth}
For a given matrix $A\in \mathbb{R}^{n\times n}$, $W\in \mathbb{R}^{n\times n}$, $r,k\in \mathbb{N}$ and $\epsilon > 0$:
\begin{enumerate}
\item with $r$ distinct rows and columns, computing $\hat{A} = UV$ as in Definition~\ref{th:wlra}, with probability $9/10$ can be done
in $O(nnz(A)+ nnz(W)+ n ^ \gamma)) + n2^{O(k^2r/\epsilon)}$, for arbitrary small $\gamma$, and hence is in FPT.

\item with $r$ distinct columns, computing $\hat{A} = UV$ as in Definition~\ref{th:wlra}, with probability $9/10$ can be done
in $O(nnz(A)+ nnz(W)+ n ^ \gamma)) + n2^{O(k^2r^2/\epsilon)}$, for arbitrary small $\gamma$, and hence is in FPT.

\item with rank $r$, computing $\hat{A} = UV$ as in Definition~\ref{th:wlra}, with probability $9/10$ can be done
in $O(n^{O(k^2r/\epsilon)}$, and hence is in FPT.

\end{enumerate}
\end{myth}

Our second example concerns with PCA as mentioned in Section~\ref{ssec:qpca}. In \cite{ppca} and \cite{apca}, Fomin et. al.
showed that for the \emph{robust PCA} and \emph{PCA with outliers} problems there are exact and randomized FPT algorithms.
Their results are based on the relation of these problems with the \emph{matrix rigidity} problem~\cite{rigid}. 
A normal PCA formulation (as in Section~\ref{ssec:qpca}) is not robust against corrupted observations.
For a data $M\in \mathbb{R}^{d\times n}$, robust PCA models outliers as additive sparse matrix $S$.
Thus we have 
a superposition of a low-rank component $L$ and a sparse component
$S$ : $M = L+S$. Now the robust PCA is as following:
\[\textit{minimize}~ rank (L)+ nnz(S)~ s.t. M = L + S\] 

On the other hand robust PCA is equivalent to matrix rigidity problem: for a matrix $M\in \mathbb{R}^{d\times n}$, suppose 
updating at most $k$ entries operation is denoted by $\textit{update}^k(M)$, then the following problem:
\[M' = \textit{update}^k(M)~s.t.~ rank (M')\leq r \]
is called matrix rigidity problem. 

\begin{myth}[\cite{rigid}]
The problem of max rigidity (robust PCA) is solvable in $poly(M) 2^{O((r+k)log(kr))}$, thus FPT, parametrized by $r+k$.
\end{myth} 

Note that this problem is NP-complete for $r\geq 0$.
Another formulation is called \emph{PCA with outliers}. The parametrized complexity of this problem in both classical and approximation setting is studied in~\cite{apca,ppca}. Here the problem is as following:

``Given a set of $n$ points in $\mathbb{R}^d$ with (unknown)
$k$ outliers, the problem is to identify the outliers so that
the remaining set of points fits best into an unknown $r$-dimensional subspace"

One can interpret this problem in the following geometrical setting:

``Given $n$ points in $\mathbb{R}^d$, represented by the rows of $A$, we
seek for a set of $k$ outliers, represented by the non-zero rows
of $N$, whose removal from $A$ leaves the remaining $n-k$
inliers as close as possible to an $r$-dimensional subspace.
The matrix $L$ contains then the orthogonal projections of
the inliers onto this subspace"

\begin{figure}[h]
\begin{center}
\includegraphics[height = 3 cm , width = 7cm]{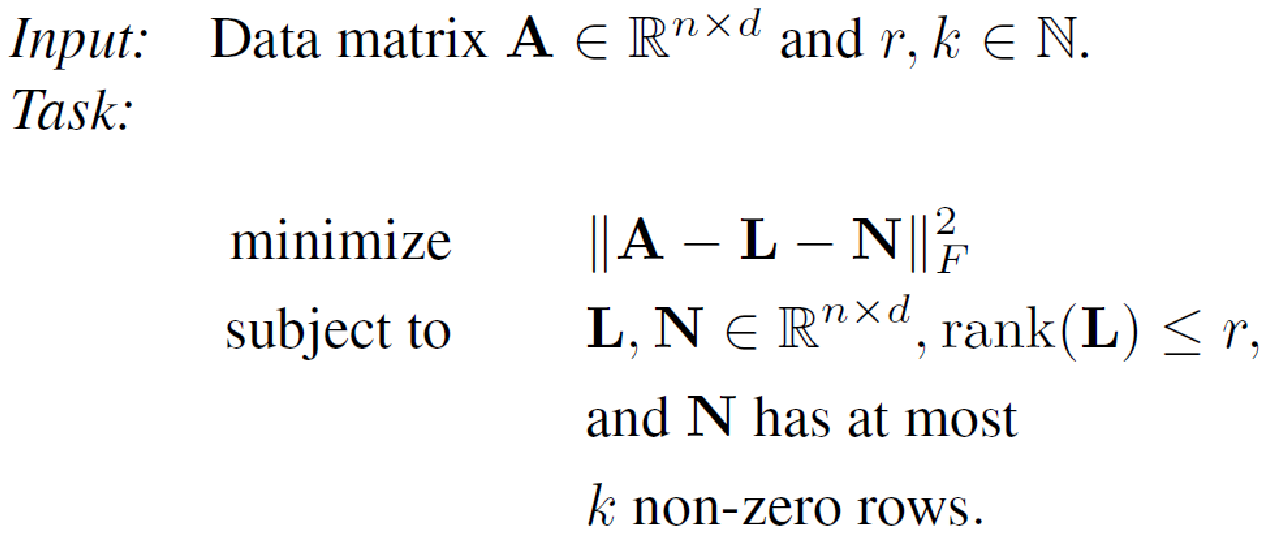} 
\end{center}
\caption{PCA with outliers}
\label{fig:pcao}
\end{figure}

Now the formal definition of the PCA with outliers is shown in Figure~\ref{fig:pcao}.

The algorithms of \cite{ppca} reduces the problem to iteration of normal PCA as in the following theorem:

\begin{myth}\label{th:rpcao}
Solving PCA with outliers is reducible to solving $|A|^{O(n^2)}$ instances of PCA. 
since only iterative algorithms for PCA and SVD exist, 
this algorithm does not solves PCA with outliers in
a fixed number of operations. However,
up to some constant precision, PCA is solvable in polynomial number of operations
\end{myth} 

Moreover by using computational complexity assertion that is called 
Exponential Time Hypothesis (ETH)~\cite{eth}, they improved Theorem~\ref{th:rpcao} to achieve fixed parameter tractability. 

Roughly ETH says that for $k\geq 3$, and $\delta_k>0$, 
$k$-SATISFIABILITY problem with $n$ variables and $m$ clauses
cannot be solved in time $O(2^{\delta_k n} (n+m)^{O(1)})$. This means
that $k$-SATISFIABILITY cannot be solved in subexponential
in $n$ time. Now the following theorem is shown in~\cite{ppca}:

\begin{myth}
For any $\epsilon \geq 1$, there is no $\epsilon$-approximation
algorithm for PCA with outliers with running time
$f(d)\, N^{o(d)}$ for any commutable function $f$ unless \textit{ETH}
fails, where $N$ is the bitsize of the input matrix.
\end{myth}

One way to circumvent ETH assumption is to consider fixed rank setting. This is investigated in~\cite{apca}, resulting in several approximation algorithm for PCA with outlier. Here we conclude this section with the following theorem, and refer the reader to ~\cite{apca} for further details.

\begin{myth}
For every $\epsilon > 0$, an $1+\epsilon$-approximate solution to PCA with outliers can be found in time 
$n^{O(\frac{r\ log\ r}{\epsilon^2})}\  d^{O(1)} $ 
\end{myth}

\begin{myconj}
Quantum inspired algorithm for recommendation system and PCA, that are based on low rank approximation and also dependent to sample and query technique, are FPT, using appropriate parametrization.
\end{myconj}

\bibliography{report}
\bibliographystyle{plain}

\end{document}